\documentclass[prc,twocolumn,floatfix,groupedaddress,nofootinbib,showpacs,preprintnumbers,
amsmath,amssymb,amsfonts,superscriptaddress] {revtex4}
\usepackage{graphicx}% Include figure files
\usepackage{dcolumn}%Align table columns on decimal point
\usepackage{mathrsfs}
\usepackage{bm}% bold math

\usepackage{graphicx}% Include figure files
\usepackage{dcolumn}% Align table columns on decimal point
\usepackage{bm}% bold math
\usepackage[usenames]{color}
\begin{document}

\title{Searching for isovector signatures in the neutron-rich oxygen
and calcium isotopes}
\author{Wei-Chia Chen}
\email{wc09c@my.fsu.edu} 
\affiliation{Department of Physics, Florida State University, Tallahassee, FL 32306} 
\author{J. Piekarewicz}
\email{jpiekarewicz@fsu.edu}
\affiliation{Department of Physics, Florida State University, Tallahassee, FL 32306}
\date{\today}
\begin{abstract}
We search for potential isovector signatures in the neutron-rich
oxygen and calcium isotopes within the framework of a relativistic
mean-field theory with an exact treatment of pairing correlations.  
To probe the isovector sector we calibrate a few relativistic density
functionals using the same isoscalar constraints but with one
differing isovector assumption. It is found that under certain conditions, 
the isotopic chain in oxygen can be made to terminate at the 
experimentally observed ${}^{24}$O isotope and in 
the case of the calcium isotopes at  ${}^{60}$Ca. To produce such 
behavior, the resulting symmetry 
energy must be soft, with predicted values for the symmetry 
energy and its slope at saturation density being
$J\!=\!(30.92\pm0.47)$\,MeV and $L\!=\!(51.0\pm1.5)$\,MeV, 
respectively. As a consequence, the neutron-skin thickness of 
${}^{208}$Pb is rather small: $R_{\rm skin}^{208}\!=\!(0.161\pm0.011)$\,fm. 
This same model---labelled \emph{``FSUGarnet''}---predicts 
$R_{1.4}\!=\!(13.1\pm0.1)$\,km for the radius of a ``canonical'' 
1.4$M_{\odot}$ neutron star, yet is also able to support a 
two-solar-mass neutron star. 
\end{abstract}
\pacs{21.60.Jz, 21.65.Cd, 21.65.Mn} 
\maketitle

Density functional theory (DFT) provides the only known tractable
framework to describe strongly interacting nuclear many-body systems
ranging from finite nuclei to neutron stars. In the spirit of DFT, the
complicated many-body effects are implicitly encoded in the parameters
of the model which in turn are determined by fitting directly to
experimental data\,\cite{Kohn:1999}. Thus, the quality of the
resultant model depends not only on the form of the functional but, in
addition, on the data used for its calibration. It is widely
recognized that the isoscalar sector of the density functional is well
constrained by available ground-state observables. This is in sharp
contrast to the isovector sector that remains poorly determined; for a
recent example see Ref.\,\cite{Chen:2014sca} and references contained
therein. Such a mismatch occurs because physical observables that are
dominated by the isoscalar sector---such as binding energies and
charge radii of many stable nuclei---have been measured with enormous
precision. Instead, data on neutron 
skins\,\cite{Abrahamyan:2012gp,Horowitz:2012tj} and neutron-star 
radii\,\cite{Ozel:2010fw,Steiner:2010fz,Suleimanov:2010th,Guillot:2013wu}, 
both highly sensitive to the isovector sector, either lack precision or are still 
open to debate. 
%Without stringent constraints, predictions of strong isovector 
%indicators must be invariably accompanied by large theoretical
%uncertainties\,\cite{Chen:2014sca}. One critical isovector observable
%is the neutron-skin thickness of heavy nuclei, a quantity that is
%strongly sensitive to the density dependence of the symmetry energy
%and one that strongly impacts many areas of nuclear physics and
%astrophysics\,\cite{Danielewicz:2002pu,Buras:2003sn,Lattimer:2004pg,
%Steiner:2004fi}.

Due to the present difficulty in obtaining accurate measurements of both
neutron skins and neutron-star radii, it seems prudent to seek alternative
isovector indicators. A fruitful arena for the search of isovector sensitivity
is pure neutron matter whose equation of state is approximately equal to 
that of symmetric nuclear matter plus the symmetry energy. The
behavior of pure neutron matter at low densities is particularly attractive 
because of its close resemblance to a resonant Fermi gas. However, although 
this has stimulated significant amount of theoretical 
activity\,\cite{Schwenk:2005ka,Gezerlis:2009iw,Vidana:2009is,Gandolfi:2009fj,
Hebeler:2009iv,Tews:2012fj,Kruger:2013kua,Gandolfi:2013baa}, one must
recognize that neutron matter remains a purely theoretical construct. A 
laboratory observable that has been identified as a strong isovector indicator is  
the electric dipole polarizability of ${}^{208}$Pb\,\cite{Reinhard:2010wz,
Piekarewicz:2010fa,Piekarewicz:2012pp,Roca-Maza:2013mla}. Indeed, 
the recent high-resolution measurement of the electric dipole 
polarizability of ${}^{208}$Pb at the Research Center for Nuclear 
Physics\,\cite{Tamii:2011pv,Poltoratska:2012nf} has provided a unique 
constraint on the density dependence of the symmetry energy and serves
as an ideal complement to measurements of the neutron skin.

Given that the symmetry energy accounts for the energy cost in departing
from equal number of protons and neutrons, one expects that the evolution 
of certain nuclear properties as one moves away from the valley of stability 
will become sensitive to the isovector nature of the interaction. For example, 
if the symmetry energy is \emph{stiff}, namely, if it increases rapidly with
density, it becomes energetically favorable to move neutrons from the 
core to the surface, resulting in a thick neutron skin\,\cite{Horowitz:2014bja}.
By the same token, a stiff symmetry energy may become small at the dilute 
nuclear surface which is of particular relevance to the valence orbitals. 
As a result, a stiff symmetry energy predicts a delay in reaching the 
neutron drip line relative to their softer counterparts\,\cite{Todd:2003xs}. 
%It is the 
%aim of this letter to explore isovector sensitivities in the dynamics of 
%neutron-rich isotopes.

Mapping the precise boundaries of the nuclear landscape has been identified 
as one of the most fundamental problems in nuclear science; 
see Refs.\,\cite{Erler:2012,Afanasjev:2013} and references contained therein. 
Although the proton drip line has been determined up to protactinium 
(atomic number $Z\!=\!91$) the neutron drip line remains unknown, except 
in the case of a few light nuclei ($Z\!\lesssim\!8$)\,\cite{Thoennessen:2004}.
A particularly dramatic example of this mismatch is the case of the fluorine 
isotopes, which have ${}^{19}$F as its only stable member.
Whereas ${}^{17}$F marks the boundary of the proton drip line, 
${}^{31}$F---with 12 neutrons away from stability---remains stable against 
strong decays. While the Coulomb repulsion is largely responsible for having 
the proton drip line just a few neutrons away from stability, the basic tenet of 
this letter is that the dynamics of the neutron drip line is highly sensitive to 
the nuclear symmetry energy.

Among the few isotopic chains with both drip-line boundaries firmly established, 
oxygen is perhaps the most intriguing one, as it provides the first clear indication 
of the emergence of a new magic number at $N\!=\!16$\,\cite{Hoffman:2009zza}. 
Whereas most mean-field calculations (both non-relativistic and relativistic) 
have predicted the stability of the ``doubly-magic" nucleus ${}^{28}$O against 
strong decays, experimental efforts have failed to find a stable isotope beyond
${}^{24}$O\,\cite{Langevin:1985,Guillemaud-Mueller:1990,Fauerbach:1996,
Hoffman:2009zza}. This \emph{oxygen anomaly} has been widely investigated
within various formulations and, to date, the most common explanation invokes 
an extra repulsion between valence neutrons generated from three-nucleon 
forces\,\cite{Otsuka:2010,Hagen:2012ox,Hergert:2013,Cipollone:2013,
Bogner:2014,Jansen:2014}. 
%%%Note, however, that a properly optimized 
%%%nucleon-nucleon interaction appears to be able to describe many aspects
%%%of nuclear structure without explicitly invoking three-nucleon 
%%%forces\,\cite{Ekstrom:2013} that two-body forces. 

The calcium isotopic chain---the next chain after oxygen with a magic number 
of protons---has also received a great deal of attention due to its 
rich subshell structure near $N\!=\!32$\,\cite{Holt:2010yb,Hagen:2012ca,
Gallant:2012as,Wienholtz:2013nya}. Particularly exciting is the recent mass 
determination of various exotic calcium isotopes---up to ${}^{54}$Ca--- at both 
TRIUMF\,\cite{Gallant:2012as} and CERN\,\cite{Wienholtz:2013nya}. Yet, 
despite these remarkable achievements, the experimental determination of 
the neutron drip line in calcium is likely years away---especially if the drip line 
is at or beyond the ``doubly-magic'' ${}^{60}$Ca.
%Owing to their prominent role, and our believe that the physics of exotic nuclei
%is sensitive to the bulk symmetry energy, we focus hereafter on the mass
%evolution of both the oxygen and calcium isotopes. This we do within the 
%framework of the relativistic mean-field (RMF) 
%theory\,\cite{Walecka:1974qa,Serot:1984ey}. 

In order to explore the sensitivity of the neutron-rich isotopes to the density 
dependence of the symmetry energy, we construct theoretical models subject 
to the same isoscalar constraints but with a single differing assumption on the 
uncertain isovector sector. In the relativistic mean-field (RMF) theory, the 
nuclear system is composed of neutrons and protons interacting via the 
exchange of various mesons and the photon. In the version of the RMF models 
employed here the interaction among the particles is described by an effective 
Lagrangian density\,\cite{Mueller:1996pm,Serot:1997xg,Horowitz:2001xj} whose 
parameters are determined by fitting model predictions to experimental data. 
In this work we employ the Lagrangian density given in Ref.\,\cite{Chen:2014sca} 
and use the same calibration scheme developed therein to find the optimal model 
parameters and their associated theoretical 
uncertainties\,\cite{Piekarewicz:2014kza}. Such a fitting protocol relies exclusively 
on genuine physical observables that can be either measured in the laboratory or 
extracted from observation. This 
approach was recently implemented in building the new \emph{FSUGold2} 
density functional\,\cite{Chen:2014sca}. The data pool of observables is 
sufficient to constrain the isoscalar sector as evinced by the very small 
associated theoretical uncertainties. However, because no inherent isovector biases 
are incorporated into the fit, FSUGold2 predicts---in accordance with most 
relativistic density functionals---a stiff symmetry energy and, as a consequence, 
a fairly thick neutron skin in ${}^{208}$Pb of 
$R_{\rm skin}^{208}\!=\!(0.287\pm 0.020)$\,fm. In an effort to explore 
the sensitivity of the isovector sector to the mass evolution along
the isotopic chains in oxygen and calcium, we now tune the density dependence 
of the symmetry energy by adding into the calibration an assumed value for the 
neutron-skin thickness of ${}^{208}$Pb. That is, we augment the calibration 
procedure by assuming values of $R_{\rm skin}^{208}\!=\!0.12$\,fm, 
$R_{\rm skin}^{208}\!=\!0.16$\,fm, and $R_{\rm skin}^{208}\!=\!0.28$\,fm---in all
three cases with an  associated error of $0.2\%$. For simplicity, the resulting
relativistic mean-field models are labeled by their assumed value of 
$R_{\rm skin}^{208}$, namely, RMF012, RMF016, and RMF028. Given that the 
data pool of observables involves doubly-magic (or semi-magic) nuclei, pairing 
correlations are not included in the calibration procedure. However, once the 
calibration is completed, we exploit our recently developed 
RMF-plus-exact-pairing (RMF+EP) approach\,\cite{Chen:2014ep} to properly 
describe the mass evolution along both isotopic chains.

%To calibrate the models, we use binding energies (of ${}^{16}$O,
%${}^{40}$Ca, ${}^{48}$Ca, ${}^{68}$Ni, ${}^{90}$Zr, ${}^{100}$Sn,
%${}^{116}$Sn, ${}^{132}$Sn, ${}^{144}$Sm, and ${}^{208}$Pb)
%\cite{Wang:2012}, charge radii (of the same nuclei except ${}^{68}$Ni
%and ${}^{100}$Sn) \cite{Angeli:2013}, giant monopole energies (of
%${}^{90}$Zr, ${}^{116}$Sn, ${}^{144}$Sm and ${}^{208}$Pb)
%\cite{Youngblood:1999}, and a maximum neutron star mass of
%2.10$M_{\odot}$ \cite{Demorest:2010bx,Antoniadis:2013}. The adopted
%errors of these observables are $0.1\%$, $0.2\%$, $1\!-\!2\%$, and
%$1\%$, respectively. It has been shown that this data pool is able to
%efficiently constrain the isoscalar sector and the associated
%uncertainties are very small \cite{Chen:2014sca}. 
%In addition, we incorporate an isovector observable, the neutron 
%skin of ${}^{208}$Pb, into the above list. In order to see the isovector 
%effects, we assume the neutron skin to be 0.12, 0.16, and 0.28 fm with 
%a $0.2\%$ error. This rather small error of the neutron skin is to ensure that
%the models fulfill the requirement, thereby having different behavior
%in the isovector sector. The resultant models are named FSUXS, FSUS,
%and FSUL in accord with the specified size of the neutron skin.
%Note that pairing is not included in the calibration since its effect on
%the fitting observables is small. After the calibration is finished,
%we exploit the approach of exact pairing (EP) \cite{Chen:2014ep} to
%deal with neutron pairing in the {\it sd} and {\it fp} shell for
%oxygen and calcium isotopes, respectively.

%%%%%  Figure 1  %%%%%
\begin{figure}[ht]
\vspace{-0.05in}
\includegraphics[width=7cm,height=7cm,angle=0]{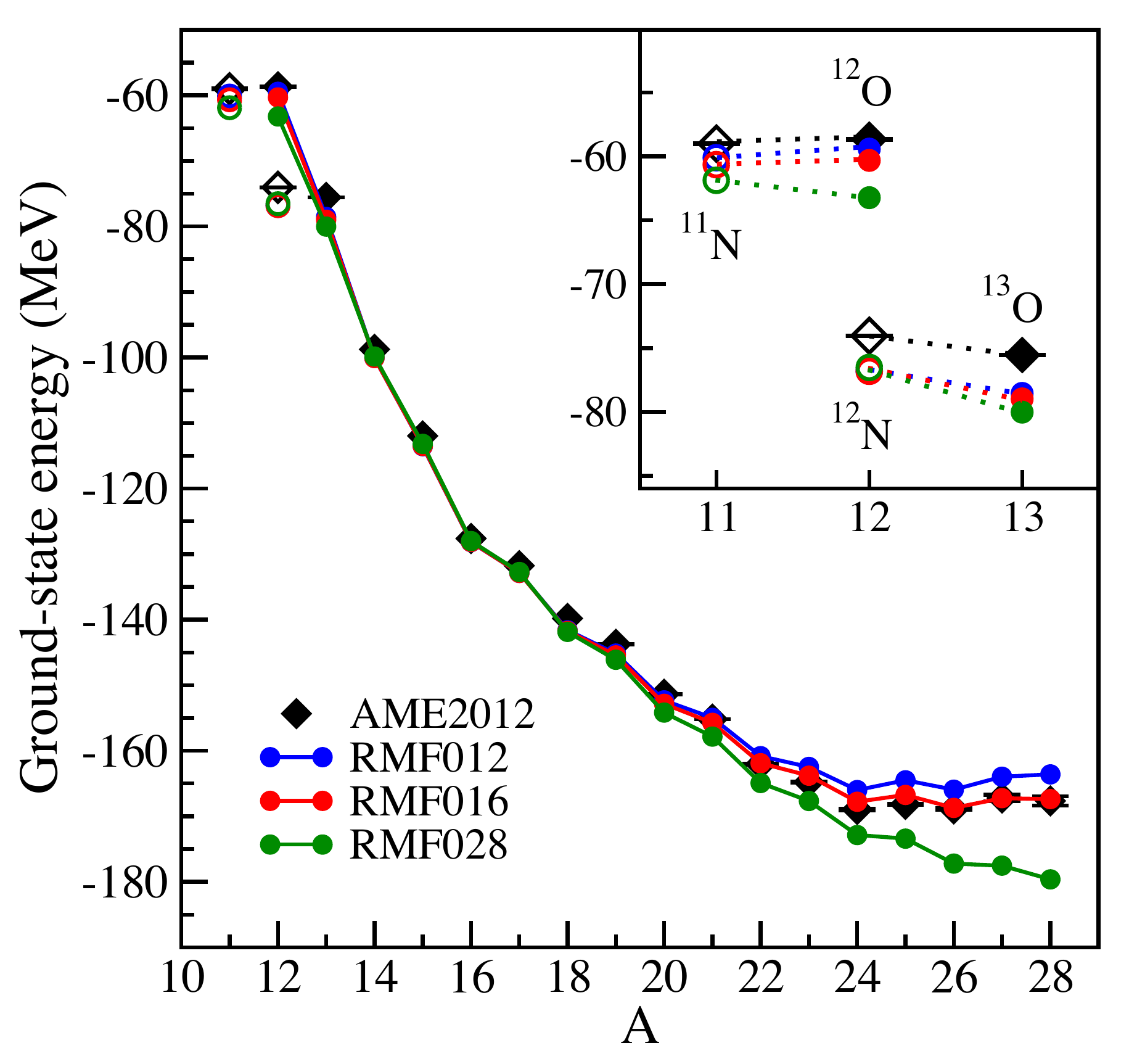}
\caption{(Color online). Evolution of the ground-state energy along the 
isotopic chain in oxygen---from ${}^{12}$O to ${}^{28}$O---as predicted 
by the three RMF models described in the text. Experimental data are 
from Ref.\,\cite{Wang:2012}.}
\label{Fig1}
\end{figure}
%%%%%%%%%%%%%%%%%%%%

We start by displaying in Fig.\,\ref{Fig1} the evolution of the ground-state energy 
along the isotopic chain in oxygen. Given that the nearly isospin-symmetric isotopes 
${}^{14-18}$O are largely insensitive to the isovector sector, the model predictions 
are almost indistinguishable from each other and are also in good agreement with 
the 2012 Atomic Mass Evaluation (AME2012)\,\cite{Wang:2012}. 
However, as the neutron-proton asymmetry is increased, the model predictions start
to differ, indicating that isovector effects are starting to play an increasingly dominant role. 
Indeed, the models display dramatic differences as the experimentally determined neutron 
drip line at ${}^{24}$O is approached. Although drip-line nuclei are undoubtedly 
sensitive to subtle dynamical effects, {\sl e.g.,} mixing to the continuum, it 
appears that the density dependence of the symmetry energy also plays a 
critical role. In particular, we find that RMF028 (with the stiffest symmetry 
energy) overbinds the neutron-rich isotopes, leading to the common, yet 
erroneous, prediction of a drip line at ${}^{28}$O. In contrast, RMF012 and 
RMF016 with a softer symmetry energy produce the necessary \emph{repulsion} 
to shift the neutron drip line to ${}^{24}$O. 
%Given that a stiff symmetry energy increases faster with density than a soft one, 
%this extra repulsion may seem contradictory at first. However, 
We must underline that such behavior is determined by the weakly-bound excess 
neutrons that reside in the nuclear surface where the density is low. Thus, it is the 
low-density component of the symmetry energy---which is \emph{larger} for a soft 
model---that dictates the physics, rather than the symmetry energy around saturation 
density. This suggests that models with a small $R_{\rm skin}^{208}$ should be the 
first ones to reach the neutron drip line\,\cite{Todd:2003xs}, precisely as seen in 
Fig.\,\ref{Fig1}. Although Coulomb effects shift the proton drip line much 
closer to stability, the imprint of the symmetry energy should also be manifest on the 
neutron-deficient side of the isotopic chain. Indeed, this appears to be the case.
As highlighted in the inset of Fig.\,\ref{Fig1}, both RMF012 and RMF016 
predict---unlike RMF028---that ${}^{12}$O is unstable against proton emission,
in agreement with experiment. Thus, as in the case of the neutron drip line,
the two softer models reach the proton drip line earlier than the stiffer one.
Note that the true ground state of the odd-odd nucleus ${}^{12}$N is a 
superposition of states with the unpaired proton and neutron being in orbitals 
that can couple to the ground-state spin of the nucleus (i.e., $J^{\pi}\!=\!1^{+}$).
However, for simplicity we approximate the ground-state energy of ${}^{12}$N
by the lowest-energy configuration. The inset on Fig.\,\ref{Fig1} seems to validate 
this approximation.

%%%%%  Figure 2a and 2b  %%%%%
\begin{figure}[ht]
\vspace{-0.05in}
\includegraphics[width=4.15cm,height=5cm,angle=0]{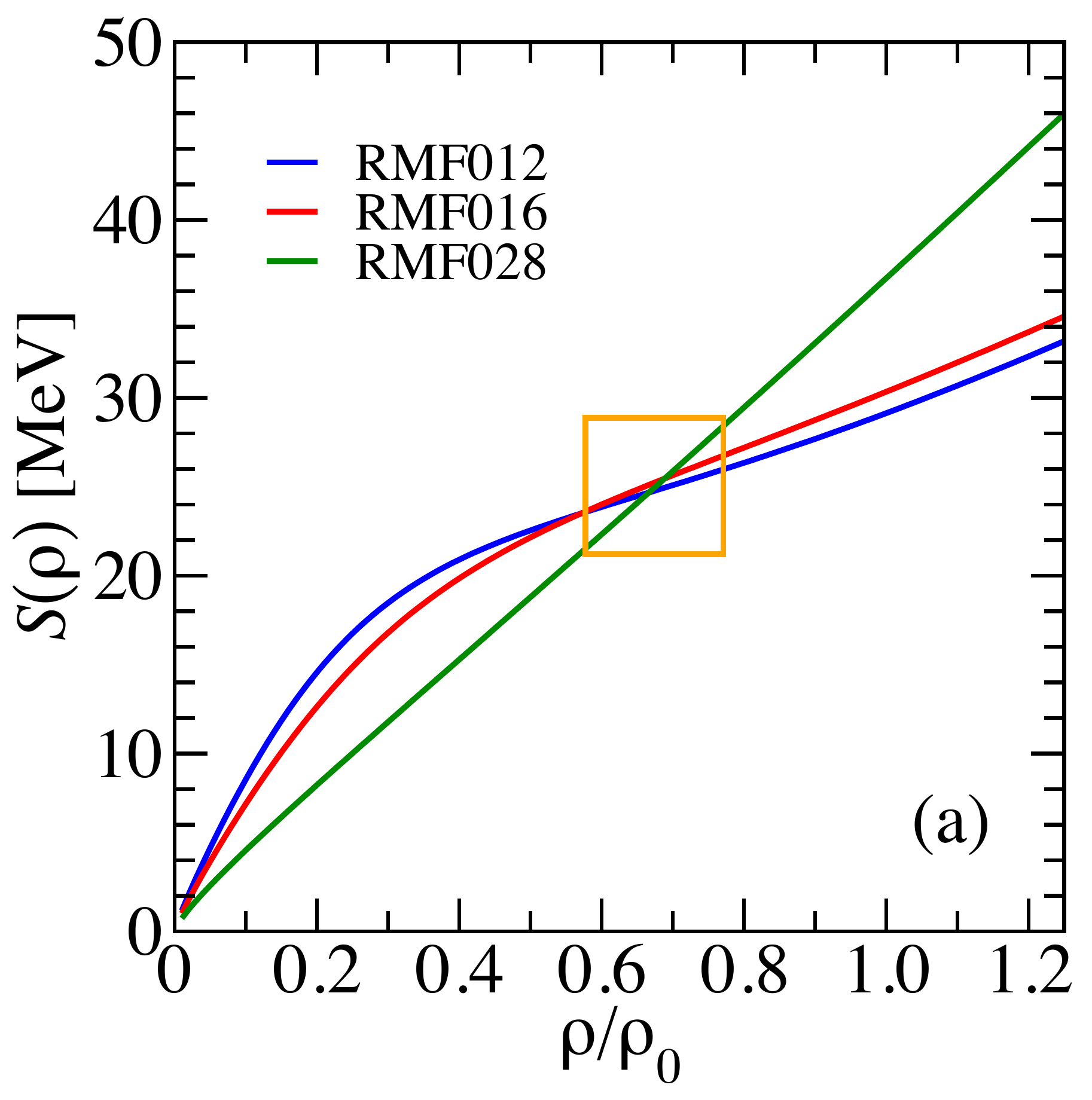}
 \hspace{2pt}
\includegraphics[width=4.15cm,height=5cm,angle=0]{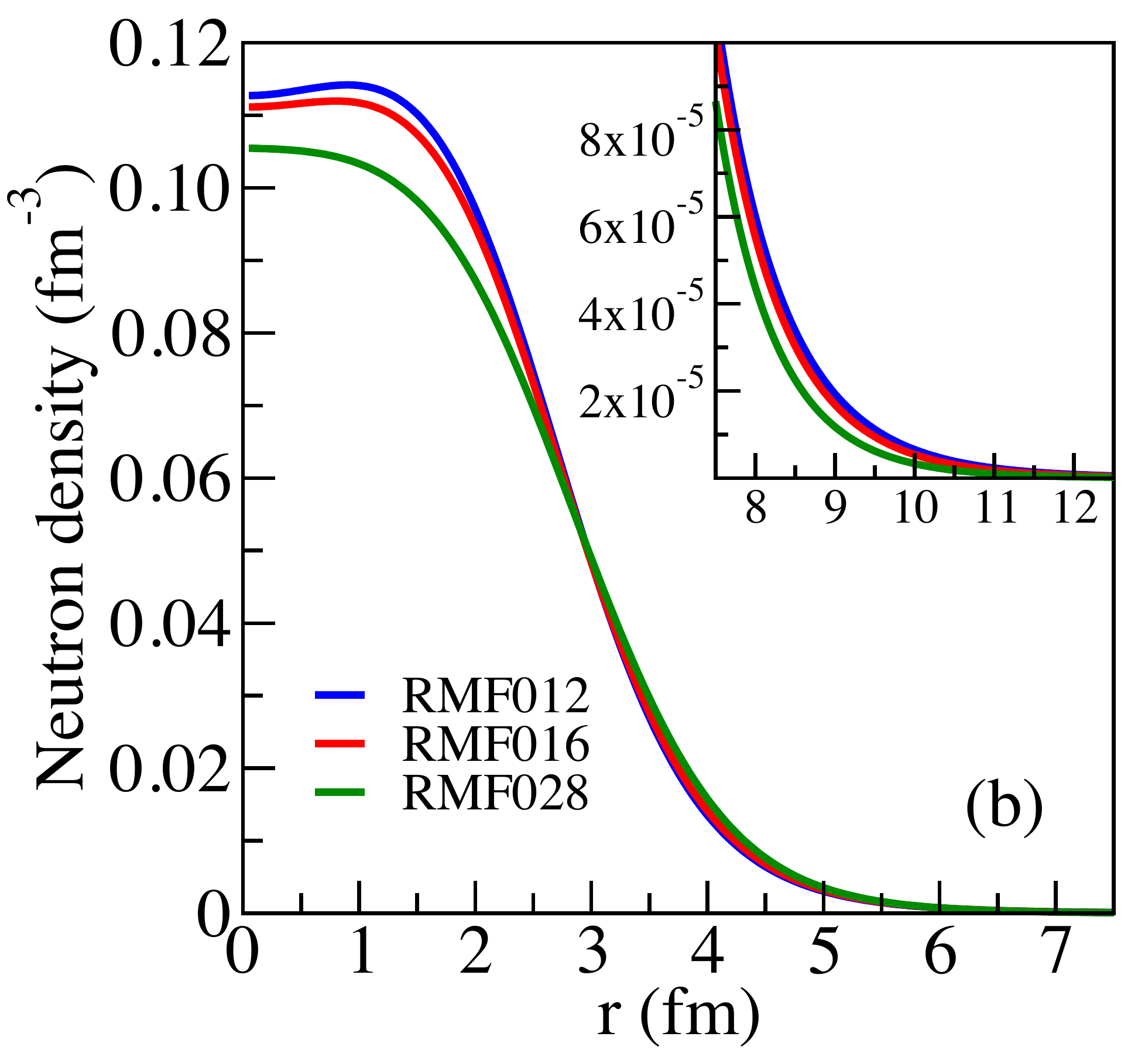}
\caption{(Color online). (a) Symmetry energy as a function of density 
in units of $\rho_{0}\!=\!0.148\,{\rm fm}^{-3}$ and (b) neutron density of
${}^{24}$O as predicted by the three RMF models discussed in the text.}
\label{Fig2}
\end{figure}
%%%%%%%%%%%%%%%%%%%%

To further validate this behavior, we display in Fig.\,\ref{Fig2}a the symmetry energy 
predicted by the three models up to a density slightly above saturation density. The 
thickness of the neutron skin in ${}^{208}$Pb is largely determined by the \emph{slope} 
of the symmetry energy at (or near) saturation density. In the case of a stiff symmetry 
energy, such as RMF028, it is energetically advantegeous to move neutrons from the 
core (where $S$ is large) to the surface (where $S$ is small), albeit at the expense of
an increase in surface tension. Thus, models with a stiff symmetry energy tend to predict 
thicker neutron skins. However, at a density of about 2/3 of saturation density, 
corresponding to a value of the symmetry energy of almost 26 MeV, all three models 
intersect each other. This well-known result emerges from the sensitivity of the binding 
energy of neutron-rich nuclei to the symmetry energy at a density that is intermediate between 
that of the core and the surface\,\cite{Farine:1978,Brown:2000,Horowitz:2001xj,Furnstahl:2002un,
Brown:2013,Zhang:2013,Horowitz:2014bja}. As a result, the symmetry energy below 
this density becomes larger for the softer models. This increase in the symmetry
energy generates the added repulsion required to shift the neutron drip line from
${}^{28}$O to ${}^{24}$O. 

%%%%%  Figure 3  %%%%%
\begin{figure}[ht]
\vspace{-0.05in}
\includegraphics[width=6.5cm,height=6.5cm,angle=0]{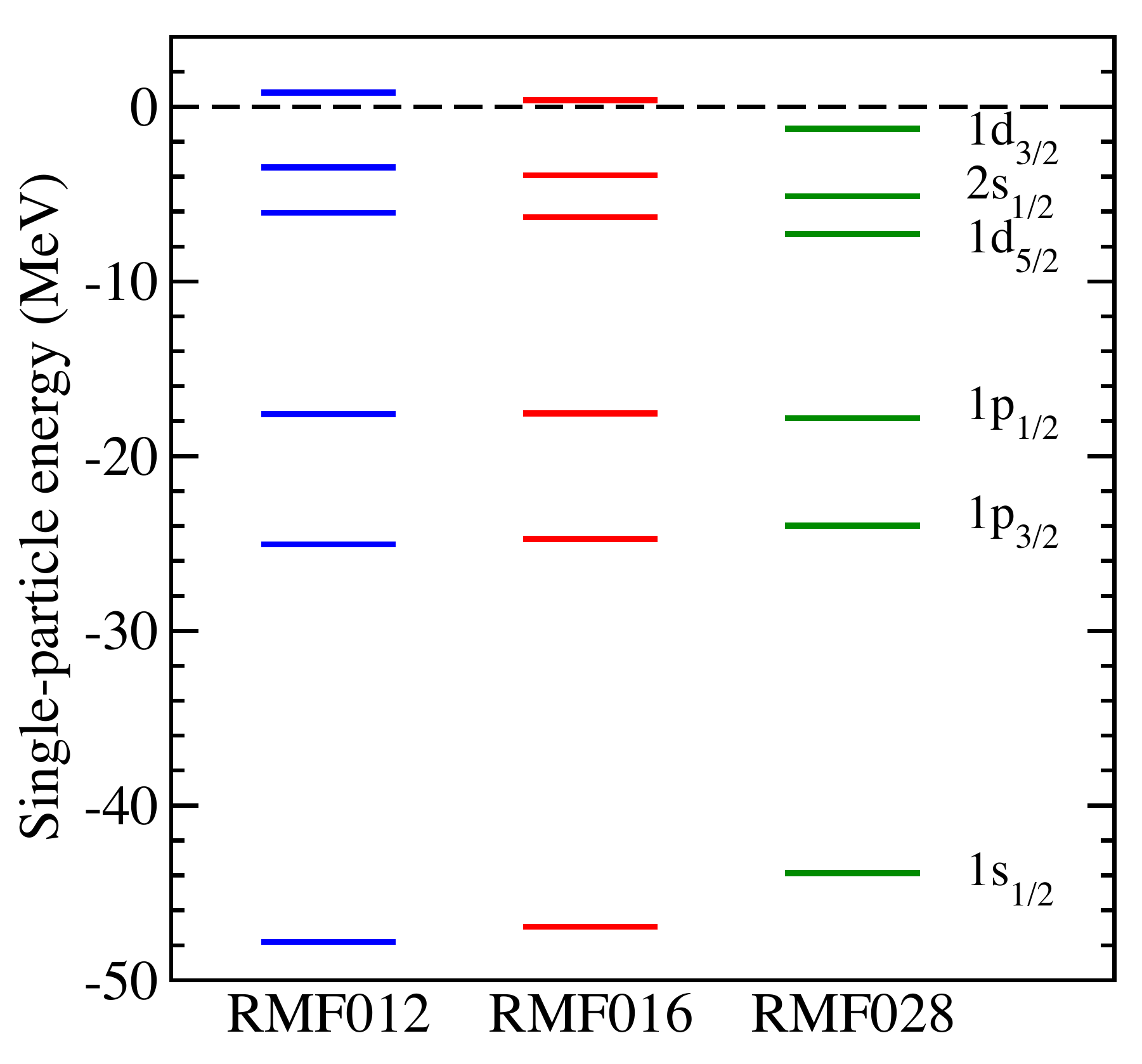}
\caption{(Color online). Single-neutron spectrum for ${}^{24}$O
as predicted by the three RMF models discussed in the text.}
\label{Fig3}
\end{figure}
%%%%%%%%%%%%%%%%%%%%

Such unique behavior of the symmetry energy leaves a distinct imprint on the neutron 
density of ${}^{24}$O; see Fig.\,\ref{Fig2}b. First, we note that for a stiff symmetry energy, 
as in the case of RMF028, more neutrons are pushed to the surface resulting in both a 
depletion of the density in the interior and a larger neutron radius. Second, at a distance 
of about 3\,fm, corresponding to a density of about 0.05\,fm$^{-3}$, the neutron density 
predicted by the RMF028 model now becomes the largest, as this is the region that 
dominates the contribution to the neutron radius. Finally, at even larger distances where 
the density is dominated by the weakly-bound valence neutrons, the density is again 
lowest for the stiffest model (see the inset in Fig.\,\ref{Fig2}b). That is, the smaller symmetry 
energy at low density of the RMF028 model yields less repulsion for the valence 
orbitals and consequently a faster falloff of the density. The single-neutron spectrum 
displayed in Fig.\,\ref{Fig3} serves to reaffirm these trends. In particular, we notice 
a ``compression" of the single-particle spectrum as the symmetry energy becomes stiffer. 
Indeed, whereas the ``core'' \emph{sp}-orbitals become less bound with increasing 
stiffness, the valence \emph{sd}-orbitals are more strongly bound. Particularly, the neutron 
$1{\rm d}_{3/2}$ orbital becomes unbound for the softer models---a critical requirement for 
the drip line in oxygen to be found at ${}^{24}$O.

%%%%%  Figure 4  %%%%%
\begin{figure}[ht]
\vspace{-0.05in}
\includegraphics[width=8.5cm,height=7cm,angle=0]{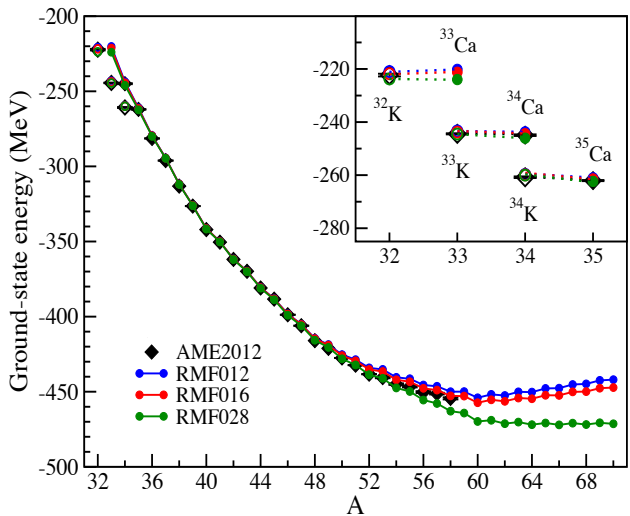}
\caption{(Color online). Evolution of the ground-state energy along the 
isotopic chain in calcium---from ${}^{33}$Ca to ${}^{70}$Ca---as predicted 
by the three RMF models described in the text. Experimental data are 
from Ref.\,\cite{Wang:2012}.}
\label{Fig4}
\end{figure}
%%%%%%%%%%%%%%%%%%%%

\begin{figure}[ht]
\vspace{-0.05in}
\includegraphics[width=6.5cm,height=6.5cm,angle=0]{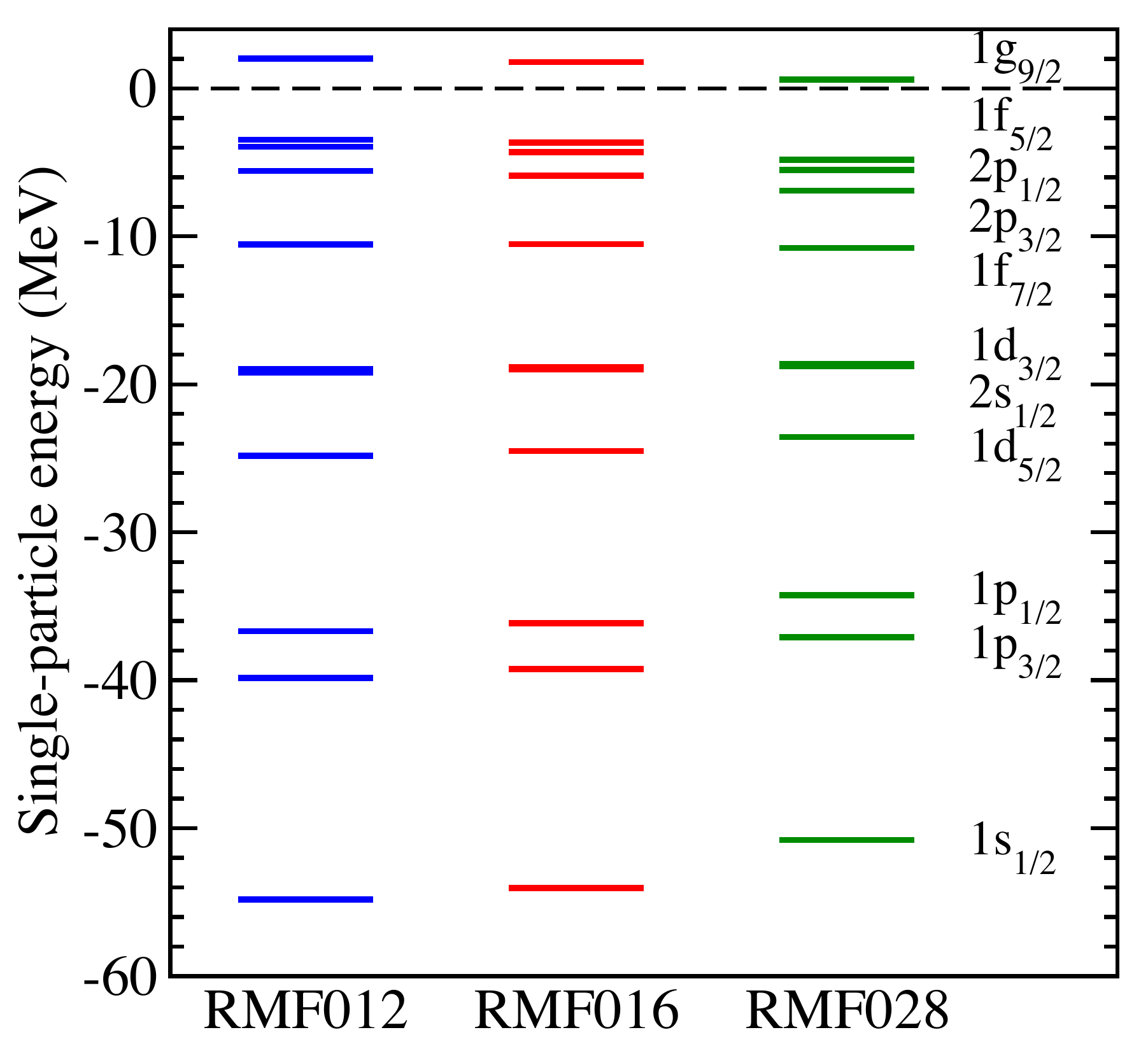}
\caption{(Color online). Single-neutron spectrum for ${}^{60}$Ca as predicted 
by the three RMF models discussed in the text. Note that the $2s_{1/2}$ and 
$1d_{3/2}$ orbitals are predicted to be nearly degenerate in all three models.}
\label{Fig5}
\end{figure}

We continue by displaying in Fig.\,\ref{Fig4} ground-state energies for calcium---the 
next isotope with a fully closed proton shell. Predictions have been made for a 
wide range of neutron-proton asymmetries starting with ${}^{33}$Ca and ending 
with the very neutron-rich ${}^{70}$Ca isotope. The calculation for the 
neutron-rich isotopes was done using the augmented $fpg_{9/2}$ valence space. 
It is found that including the $1g_{9/2}$ orbital enhances the binding energy in the 
${}^{40-60}$Ca region bringing the predictions from both RMF012 and RMF016 
into closer agreement with experiment.
Also shown in the figure are experimental data from the latest AME2012 
compilation\,\cite{Wang:2012}. Note that the AME2012 results quoted for ${}^{53}$Ca 
and beyond were ``derived not from purely experimental data''\,\cite{Wang:2012}. 
Contrary to the isotopic chain in oxygen where the neutron drip line has been firmly 
established, the experimental data show no evidence that the neutron drip line is within 
reach. Given that all three models were calibrated using ground-state energies for both 
${}^{40}$Ca and ${}^{48}$Ca, it is not surprising that the agreement among 
them---and with experiment---is very good. However, beyond ${}^{48}$Ca where 
isovector effects start to play a critical role, significant differences emerge. In 
particular, and fully consistent with the results obtained along the isotopic chain 
in oxygen, the stiff RMF028 model predicts an overbinding that is inconsistent 
with experiment. 
%Although the figure also suggests a slight underbinding of the two
%softer models, the predicted trend is consistent with experiment. 
Although all three models agree that ${}^{60}$Ca is particle bound, a 
subtle odd-even staggering emerges thereafter. Nevertheless, upon closer examination 
we found that for RMF012 and RMF016 the neutron drip line is reached at ${}^{60}$Ca. 
On the other hand, the very flat plateau displayed by RMF028 makes it difficult to 
identify the exact location of the neutron drip line.
This observation is also supported by the single-neutron energies of ${}^{60}$Ca
displayed in Fig.\,\ref{Fig5}. Indeed, compared to its softer counterparts, the barely
unbound $1g_{9/2}$ orbital in RMF028 makes the identification of the drip line ambiguous. 
We stress that a more accurate description of the neutron drip line remains a serious 
theoretical challenge. For example, whereas Holt and collaborators\,\cite{Holt:2010yb} 
predict---like we do---that the neutron drip line will be reached at or beyond ${}^{60}$Ca, 
Hagen \emph{et al.} find ${}^{60}$Ca to be particle unbound relative to 
${}^{56}$Ca\,\cite{Hagen:2012ca}. However, while the results of Holt \emph{et al.} 
depend critically on the role of three-nucleon forces, Ekstr\"om and collaborators 
have recently found that a properly optimized chiral nucleon-nucleon interaction can
describe many aspects of nuclear structure without explicitly invoking three-nucleon 
forces\,\cite{Ekstrom:2013}.

Finally, we turn to the neutron-deficient side of the calcium isotopes. As 
shown in Fig.\,\ref{Fig4} and highlighted in the inset, models with a soft symmetry
energy reach both drip lines earlier than their stiffer counterparts. Indeed, whereas 
the proton drip line in both RMF012 and RMF016 can be placed at ${}^{34}$Ca in 
agreement with experiment, RMF028 predicts its location at or beyond ${}^{33}$Ca. 
(Again, for the odd-odd nuclei, ${}^{32}$K and ${}^{34}$K, we used the same 
approximation as for ${}^{12}$N.)

%%%%%  Figure 5  %%%%%
%%%%%%%%%%%%%%%%%%%%

The results obtained so far suggest that by adopting certain reasonable 
assumptions one can reproduce the observed experimental trends in the 
isotopic chains of both oxygen and calcium. In our particular case, such 
a ``plausible assumption'' implies the adoption of an ad-hoc value of 
$R_{\rm skin}^{208}$. This finding is significant as it has been previously 
shown that calibrating RMF functionals by relying exclusively on 
well-measured physical observables invariably results in the prediction of fairly 
large neutron skins\,\cite{Fattoyev:2013yaa,Chen:2014sca,Piekarewicz:2014kza}. 
%Thus, ground-state energies of neutron-rich nuclei along isotopic chains 
%become---together with the neutron-skin thickness and the electric dipole 
%polarizability---a physical observable that is highly sensitive to isovector effects.
Instead, one immediate consequence of the present analysis is that the 
neutron-skin thickness of neutron-rich nuclei can not be overly large. Indeed, 
the model that can best reproduce ground-state energies along the oxygen 
and calcium isotopic chains is RMF016---a model that henceforth will be 
referred to as \emph{``FSUGarnet''}. This model predicts
$R_{\rm skin}^{208}\!=\!(0.161\pm0.011)$\,fm. 
%Although at the lower end of the PREX measurement of 
%$R_{\rm skin}^{208}\!=\!(0.33^{+0.16}_{-0.18})$\,fm\,\cite{Abrahamyan:2012gp,
%Horowitz:2012tj}, our prediction remains consistent within the wide experimental range. 
In turn, the strong correlation between $R_{\rm skin}^{208}$ and 
neutron-star radii leads to the following prediction for the radius of a 1.4$M_{\odot}$ 
neutron star: $R_{1.4}\!=\!(13.1\pm0.1)$\,km. Note that although the symmetry 
energy is relatively soft, the overall equation of state is stiff enough to support 
a two-solar-mass neutron star\,\cite{Demorest:2010bx,Antoniadis:2013pzd}. 
Indeed, the maximum neutron-star mass supported by FSUGarnet is 
$M_{\rm max}\!=\!(2.07\pm0.02)\,M_{\odot}$. Finally, given that no property of 
infinite nuclear matter was incorporated into the fit, the symmetry energy 
$J\!=\!(30.92\pm0.47$)\,MeV and its slope $L\!=\!(51.0\pm1.5$)\,MeV at 
saturation density represent legitimate model predictions. 

Although all our results were obtained from the calibration of a relativistic density 
functional constrained exclusively from experimental and observational data---plus 
a critical assumption on $R_{\rm skin}^{208}$---it is instructive to compare them
against the predictions from various other analyses. In the particular case of the 
neutron-skin thickness of ${}^{208}$Pb, it falls safely within the 
$R_{\rm skin}^{208}\!=\!(0.14\!-\!0.23)$\,fm range suggested by a myriad of 
different analyses\,\cite{Carbone:2010az,Hebeler:2010,Tamii:2011pv,Moller:2012,
Steiner:2012,Tsang:2012,Lattimer:2012,Hebeler:2013,Lattimer:2013}. In regard to 
the symmetry energy and its slope at saturation density, many of these same 
publications are consistent with the predictions from FSUGarnet. This is not overly 
surprising given that the value of the symmetry energy $J$ is largely constrained 
by nuclear masses and its slope $L$ by the value of $R_{\rm skin}^{208}$. 

In summary, we have explored sensitivity to isovector effects in the neutron-rich 
oxygen and calcium isotopes by calibrating RMF models with the same isoscalar 
constraints but with a single differing assumption on the isovector sector: the 
neutron-skin thickness of ${}^{208}$Pb. We found that in these neutron-rich isotopes 
isovector effects play a critical role in reproducing the correct experimental trends 
along both isotopic chains. In particular, FSUGarnet---a newly calibrated relativistic 
density functional---displays a soft symmetry energy that can provide the extra 
repulsion required to terminate the oxygen chain at ${}^{24}$O and 
predicts the neutron drip line in calcium to be reached at ${}^{60}$Ca. The same 
isovector trends were also found on the neutron-deficient side of both isotopic 
chains. Indeed, FSUGarnet predicts the proton drip line in oxygen and calcium 
to be reached at ${}^{13}$O and ${}^{34}$Ca, in agreement with experiment.
Although we have established the critical role that the symmetry energy plays 
in the delineation of the drip lines, we recognize that our results may be model
dependent (see Ref.\,\cite{Co:2012}). Yet, we are confident that our findings are 
of sufficient interest to motivate alternative studies with other classes of density 
functionals.

This material is based upon work supported by the U.S. Department of 
Energy Office of Science, Office of Nuclear Physics under Award Number 
DE-FD05-92ER40750.

%\bibliography{./ReferencesJP}
\bibliography{./RMFIsotopes.bbl}

\end{document}